\documentclass{article}
\usepackage{amsmath}

\usepackage{aip}

\input{tcilatex}

\begin{document}

\title{A new approach to strongly correlated disorder }
\author{Z. Shemer and V. Halpern*}
\date{ }
\maketitle

\begin{abstract}
Problems involving disordered systems are usually analyzed for systems with
random disorder. However, there are many systems in which the main disorder
involves clusters with correlated differences between their properties and
those of the average system. A new approximation, the average trace
approximation, is proposed for calculating the diagonal elements of the
Green function, and hence the density of states, in such systems. As an
example, application of the method to a simple cubic array of harmonic
oscillators shows that correlation in the disorder leads to a peak in the
low frequency density of states, a result confirmed by computer simulations.
\end{abstract}

There are many problems in the physics of condensed matter that involve the
behavior of disordered systems. The simplest type of disorder involves
fluctuations in the diagonal elements of the system's Hamiltonian. For
electronic systems, such problems correspond to fluctuations in the
electron's self energy, while for vibrational problems it corresponds to
isotopic impurities, i.e. fluctuations in the masses of the atoms. The most
powerful method to study the effect of such disorder is the coherent
potential approximation (CPA) \cite{CPA}. Another type of disordered system
that has been widely studied is that of off-diagonal disorder in the
elements of the Hamiltonian between adjacent sites. Examples of this for
electrons include random hopping rates between different sites \cite{Odagaki}
and the migration of localized electronic excitations among guest molecules
in a host \cite{Webman}. The corresponding problem for vibrations is that of
bond disorder, which is modelled by systems of harmonic oscillators with
random distributions of spring constants \cite{Schir-PRL} \ All these
authors use a coherent medium approximation (CMA) that is essentially
equivalent to the CPA. Spin glasses are an example of a more complicated
random system, which we will not consider in this paper, since they are not
usually treated in terms of Green functions.

In all the above system, it is usually assumed that the disorder is random.
Even when correlated disorder is considered, it is often treated in terms of
a probability distribution of the correlation decaying with distance between
the sites \cite{Dunlap}. This is not the type of correlation that is
relevant to specific defect clusters involving several sites or bonds with
correlated properties different from those of the surrounding medium. Such
defects can easily occur, for instance with the bonds to atoms with
irregular coordination numbers and the electronic states associated with
them. Other examples include clusters of molecules in liquids, and atoms or
electronic states in the region of grain boundaries in crystals or in
regions of increased or decreased density in glasses and fluids. For
electronic systems, the homomorphic coherent cluster approximation (HCPA) %
\cite{HCPA} was proposed (which actually was the original basis for the CMA)
to treat such problems, and had some success. However, there are grave
practical difficulties in using it in more than one dimension for clusters
containing more than two sites, since it is difficult to find clusters of
identical shape (homomorphic) that span the whole system and for which
calculations are practicable. For instance, the smallest such cluster for a
simple cubic lattice contains seven sites, the central one and its six
nearest neighbors. In addition, a fundamental objection has been raised to
the use of the HCPA \cite{HCPA-obj}, namely that the sites on the edge of
the cluster are regarded partly as belonging to the cluster and partly to
the average surrounding medium, so that it is far from clear when use of the
HCPA is justified. In this paper, we present a new method for calculating
the density of states in systems containing a low density of clusters with
correlated disorder, which is based on the T-matrix formalism for the
calculation of Green functions and a new approximation to the diagonal
elements of the Green function which we call the average trace approximation
(AVTA). As an example of the application of this method, we consider its
application to a simple cubic array of harmonic oscillators containing
defects for which all the spring constants are weak. For this system, we
find a low frequency peak in the density of states which is not present for
the same amount of random disorder, and our results (and hence the accuracy
and usefulness of the AVTA) are confirmed by computer simulations.

The systems that we consider have a Hamiltonian of the form 
\begin{equation}
H=H_{0}+H_{p}
\end{equation}%
where $H_{0}$ is a Hamiltonian having a known Green function $G_{0}$ and the
perturbation (not necessarily small) $H_{p}$ can be expressed in the
appropriate basis set as the sum of Hamiltonians $H_{\alpha },\quad \alpha
=1...M$ associated with $M$ disjoint clusters $\ $For convenience, we will
consider the most common case, in which $H_{0}$ is the Hamiltonian of a
crystal and we use the site representation, in which we denote a site by $%
\mathbf{n}$, so \ that an operator $F$ can be written as $\sum_{\mathbf{m}%
}\sum_{\mathbf{n}}|\mathbf{m}>F<\mathbf{n}|$. Our aim is to calculate the
density of states $g(E)$ in the system from the trace of its Green function %
\cite{Economou}, 
\begin{equation}
g(E)=\mp \frac{1}{\pi }\func{Im}\{Tr[G(E\pm i\epsilon )]\}\text{.}
\end{equation}%
For the calculation of $G$, we use the T-matrix formalism, and write 
\begin{equation}
G(z)=G_{0}(z)+G_{0}(z)T(z)G_{0}(z),
\end{equation}%
where 
\begin{equation}
T(z)=H_{p}[I-G_{0}(z)H_{p}]^{-1}.
\end{equation}%
For random disorder, the above formalism is usually used either in the
average T-matrix approximation (ATA), where one writes $%
<G(z)>=G_{0}(z)+G_{0}(z)<T(z)>G_{0}(z)$,$\ $or in the CPA, in which one
finds the value of $z$ for which $<T(z)>=0$. However, even the extension of
the ATA to clusters of more than two sites involves all the above-mentioned
problems of the HCPA. In addition, even for two sites there are problems in
defining the operator $P$ required to solve equation (7), as described below
after equation (12). In this paper, we propose an entirely different method
for calculating the diagonal elements of $G$, and hence the density of
states.

While a general operator $F$ has non-zero matrix elements between any pair
of sites, the Hamiltonian $H_{p}$ of the perturbation is the sum of
Hamiltonians $H_{\alpha }$ each of which is associated with a cluster
containing only a few sites. In that case, if a cluster centred around $%
\mathbf{n}_{\alpha }$ contains only the site $\mathbf{n}_{\alpha }$ and some
adjacent sites $\mathbf{n}_{\alpha }+\mathbf{a}_{k},\quad k=1..R$, it is
convenient to introduce the row vector $|\alpha >=(|\mathbf{n}_{\alpha }>,|%
\mathbf{n}_{\alpha }\mathbf{+a}_{1}>,...,|\mathbf{n}_{\alpha }\mathbf{+a}%
_{R}>)$, as proposed by Economou \cite{Economou} In order to present our
method, we consider first the conventional case where each cluster consists
of only a single site, so that two clusters $\alpha =1,2$ contain only $|%
\mathbf{n}_{1}>$ and $|\mathbf{n}_{2}>$ respectively, and write $G(z;\mathbf{%
n}_{1},\mathbf{n}_{2}\mathbf{)}$ = $<\mathbf{n}_{1}|G(z)|\mathbf{n}_{2}>$,
and similarly $T(z;\mathbf{n}_{1},\mathbf{n}_{2}\mathbf{)}$ = $<\mathbf{n}%
_{1}|T(z)|\mathbf{n}_{2}>$. In this notation, we can write 
\begin{equation}
H_{1}=\sum_{\alpha =1}^{M}H_{\alpha }=\sum_{\alpha =1}^{M}|\mathbf{n}%
_{\alpha }>V_{\alpha }<\mathbf{n}_{\alpha }|.
\end{equation}%
where $V_{\alpha }$ is the matrix (in this case a scalar) describing the
perturbation at site $\mathbf{n}_{\alpha }$. For a system containing only a
single defect at site $\mathbf{n}_{\alpha }$, it readily follows from
equations (4) and (5) that the corresponding operator $T_{\alpha }$ is just $%
T_{\alpha }=|\mathbf{n}_{\alpha }>t_{\alpha }<\mathbf{n}_{\alpha }|$, where $%
t_{\alpha }$ is a scalar, 
\begin{equation}
t_{\alpha }=V_{\alpha }[1-<\mathbf{n}_{\alpha }|G_{0}|\mathbf{n}_{\alpha }>%
\mathbf{V}_{\alpha }]^{-1}.
\end{equation}%
If there are a number of defects, one can write 
\begin{equation}
T=\underset{\alpha }{\tsum }Q_{\alpha }=\underset{\alpha }{\tsum }[T_{\alpha
}+T_{\alpha }G_{0}\underset{\beta \neq \alpha }{\tsum }Q_{\beta }],
\end{equation}%
where $Q_{\alpha }$\ is the contribution to the scattering matrix $T$ from
the defect at $\mathbf{n}_{\alpha }$ in the presence of all the other
defects. Iteration of this equation leads to the standard equation 
\begin{equation}
Q_{\alpha }=\underset{\beta \neq \alpha }{\tsum }\,\underset{\gamma \neq
\beta }{\tsum {\tiny \cdot \cdot }}T_{\alpha }+T_{\alpha }G_{0}T_{\beta
}+T_{\alpha }G_{0}T_{\beta }G_{0}T_{\gamma }+...
\end{equation}

These equations can be generalized to defects containing more than one site
by replacing the scalars $|\mathbf{n}_{\alpha }>$ with the vectors $|\alpha >
$\ throughout, in which case $V_{\alpha }$, $G_{\alpha }=<\alpha
|G_{0}|\alpha > $ and $t_{\alpha }$ will be square matrices. In order to
show this, we consider a specific cluster $\alpha $, and write $\mathbf{m}%
_{k}=\mathbf{n}_{\alpha }+\mathbf{a}_{k}$. Then, for instance 
\begin{equation}
H_{\alpha }=|\alpha >V_{\alpha }<\alpha |=\sum_{k,l}|\mathbf{m}_{k}>V_{kl}<%
\mathbf{m}_{l}|,
\end{equation}%
and since $<\mathbf{r}|\mathbf{s}>=\delta _{\mathbf{r},\mathbf{s}}$ it
follows that $<\mathbf{r|}H_{\alpha }|\mathbf{s}>=\delta _{\mathbf{r},%
\mathbf{m}_{k}}\delta _{\mathbf{s},\mathbf{m}_{l}}V_{kl}$, and so that $%
V_{kl}=<\mathbf{m}_{k}|H_{\alpha }|\mathbf{m}_{l}>$ is the element $(k,l)$
of the matrix $V_{\alpha }=<\alpha |H_{\alpha }|\alpha >$. Similar
definitions apply to the matrices $G_{\alpha }=<\alpha |G_{0}|\alpha >$ and $%
t_{\alpha }$, and it can easily be shown that the equivalent of equation (6)
for our matrices is 
\begin{equation}
t_{\alpha }=V_{\alpha }[I-G_{\alpha }\mathbf{V}_{\alpha }]^{-1}
\end{equation}%
Thus, in general we can write 
\begin{equation}
T=\sum_{\alpha }\underset{\beta \neq \alpha }{\tsum }\underset{\gamma \neq
\beta }{\tsum {\tiny \cdot \cdot }}T_{\alpha }+T_{\alpha }G_{0}T_{\beta
}+T_{\alpha }G_{0}T_{\beta }G_{0}T_{\gamma }+...,
\end{equation}%
where now 
\begin{equation}
T_{\alpha }=|\alpha >t_{\alpha }<\alpha |=\sum_{k,l}t_{kl}|\mathbf{m}_{k}><%
\mathbf{m}_{l}|.
\end{equation}%
However, while for single site clusters equation (7) can be solved by
defining an operator $P=G_{0}-<\mathbf{n}_{\alpha }|G_{0}|\mathbf{n}_{\alpha
}>$, this method does not apply to solving equation (11) for clusters
containing more than one site. Instead, we note that for the density of
states we only need the trace of $G$ and so of $G_{0}$ and of $G_{0}TG_{0}$.
In order to calculate the latter, we introduce the average trace
approximation (AVTA), in which we write 
\begin{equation}
\langle Tr\{G_{0}TG_{0}\}\rangle \approx \langle Tr\{G_{0}(\sum_{\alpha
=1}^{M}T_{\alpha })G_{0}\}\rangle =M\langle Tr\{G_{0}|\alpha >t_{\alpha
}<\alpha |G_{0}\}\rangle ,
\end{equation}%
where $\langle .\rangle $ denotes the mean value and $M$ is the number of
defect clusters. In the AVTA it is essential to take the trace before
averaging, since (in contrast to the ATA and CPA) we do not average out the
defects over the whole system and calculate $\langle G\rangle $ for the
periodic average system. The justification of the AVTA for small defect
densities is that the first term which we are ignoring in $Tr\{G_{0}TG_{0}\}$
is $Tr\{G_{0}T_{\alpha }G_{0}T_{\beta }G_{0}\}$ where $\beta \neq \alpha $.
On taking the trace in the site representation, we write 
\[
Tr\{G_{0}T_{\alpha }G_{0}T_{\beta }G_{0}\}=\sum_{\mathbf{n}}<\mathbf{n}%
|G_{0}|\alpha >t_{\alpha }<\alpha |G_{0}|\beta >t_{\beta }<\beta |G_{o}|%
\mathbf{n}>. 
\]
Each term of this sum involves the matrix elements of $G_{0}$ between sites
belonging to two different defects multiplied by the matrix elements of $%
G_{0}$ between a site $\mathbf{n}$ and one site from each defect. These
terms will be small if the defects are far apart, as they will be for low
defect densities. In terms of Feynman diagrams, the first terms that we are
ignoring correspond to propagation from an arbitrary site to one defect,
from that to the second defect, and then back to the original site. If we
were interested in non-diagonal elements of $G$, for some of which a path
starts at a site close to cluster $\alpha $ and ends at a site close to
cluster $\beta $, this approximation would not be justified.

For the calculation of $Tr\{G_{0}|T_{\alpha }|G_{0}\}$ for a given cluster $%
\alpha $, we write $G_{0}(z)$ in the form $G_{0}(z)=\sum_{\mathbf{k}}|%
\mathbf{k}><\mathbf{k}|/(z-E_{\mathbf{k}})$ and use equation (12) for $%
T_{\alpha }$, while $<\mathbf{k|n}>=\exp (-i\mathbf{k}\cdot \mathbf{n})$. It
then follows that 
\begin{equation}
Tr\{G_{0}(z)|T_{\alpha }(z)|G_{0}(z)\}=\frac{M}{N}\sum_{l,m}\sum_{\mathbf{k}%
}t_{lm}\frac{\exp [i\mathbf{k}\cdot (\mathbf{a}_{m}-\mathbf{a}_{l})]}{(z-E_{%
\mathbf{k}})^{2}},
\end{equation}%
which depends only on the distances $\mathbf{a}_{m}-\mathbf{a}_{l}$ between
pairs of sites in a cluster. In order to evaluate the trace in terms of the
known expressions for $G_{0}(z;\mathbf{a}_{l},\mathbf{a}_{m})$, we write
equation (14) in the form%
\begin{equation}
Tr\{G_{0}(z)|T_{\alpha }(z)|G_{0}(z)\}=-M\frac{\partial }{\partial z}%
\sum_{l,m}t_{lm}G_{0}(z;\mathbf{a}_{l},\mathbf{a}_{m}).
\end{equation}

In order to demonstrate the usefulness and accuracy of the AVTA, we now
apply it to a specific system, and compare its results with those of
computer simulations. The model system that we consider is a simple cubic
lattice of coupled harmonic oscillators with spring constants $K$. For the
sake of simplicity we chose a binary distribution, with a small fraction of
oscillators having a spring constant $K_{1}$ (which we refer to as weak
springs) much less than the spring constant $K_{0}$ of most of the
oscillators. Extensive results for a random distribution of weak springs in
this system have been presented by Schirmacher and Diezemann.\cite{Schir},
who found that a concentration of 10\% of weak springs had a negligible
effect on the density of vibrational states. In contrast to their model, we
considered systems in which the weak springs occur in clusters, with the
same type of cluster in each defect. The defects that we consider are those
in which a site is linked to its six neighbors by weak springs, with $%
K_{1}/K_{0}=0.1$. The results for more general systems of this type, in
which a site is linked to its six neighbors by $z$ weak springs with spring
constant $K_{1}$ and $6-z$ ordinary springs with spring constant $K_{0}$, as
well as the results for other values of $K_{1}/K_{0}$, will be reported
elsewhere \ The concentration $c=M/N$ of these defects was chosen to be
1.5\%, so that the total concentration of weak springs was 9\%. Since the
clusters are identical, it follows from equation (15) that they each give
the same contribution to $Tr\{G_{0}T_{\alpha }G_{0}\}$. Thus it is
sufficient to calculate $Tr\{G_{0}T_{\alpha }G_{0}\}$ for one such cluster,
so that 
\begin{equation}
Tr\{G\}=Tr\{G_{0}\}+cNTr\{G_{0}T_{\alpha }G_{0}\},
\end{equation}%
and for the density of states per site 
\begin{equation}
\frac{1}{N}Tr\{G\}=G_{0}(z;\mathbf{0},\mathbf{0})-c\frac{\partial }{\partial
z}\sum_{l,m}t_{lm}G_{0}(z;\mathbf{a}_{l},\mathbf{a}_{m}).
\end{equation}%
In order to check the accuracy of the AVTA, we also performed a number of
simulations on cubes with 13-16 oscillators on each side, and took a
smoothed average of the density of states obtained for these systems.

The results of our calculations are shown in figure 1, in which we compare
the densities of states in the periodic system (no defects) with that in the
system with a concentration of 1.5\% of these defects. We note first that in
the periodic system the density of states $g(\omega )$ is fairly smooth,
with the well-known van-Hove singularities, where there is a discontinuity
in the gradient of $g(\omega )$, at $\omega =2$ and $\omega =\sqrt{8}.$ On
the other hand, both the calculations and the simulations show a large peak
in $g(\omega )$ around $\omega =0.75$. This peak is not present for this
density (9\%) of randomly placed weak springs, as found by \cite{Schir} \
and also in our own calculations and simulations for uncorrelated weak
springs. It is very reminiscent of the peak in the density of low frequency
vibrations found in glasses, where it is known as the Boson peak \cite%
{Schir-PRL}.\ The difference in the heights of the maxima between the
calculations and simulations are associated with the smoothing that has to
be performed for the latter, and the areas under the peaks, which correspond
to the total extra density of states in this region, are very similar. The
slight difference in the positions of the peak may also be associated with
this smoothing. However, it might just be a genuine effect arising from
interactions between the defect clusters which are ignored in the AVTA but
could occur in the simulated systems (where noise places a lower limit to
the defect concentrations that can be studied). Incidentally, in the
calculated density of states, there are slight problems at the frequencies
of the Van-Hove singularities, as can be seen in figure 1, because of our
use of equation (17), since the derivative of $G_{0}(z)$ is discontinuous at
these points.

The main conclusions from this paper are that the presence of defect
clusters involving correlated disorder extending over several sites can have
an appreciable effect on the density of states. The average trace
approximation (AVTA) presented in this paper provides a simple (at least in
principle) and accurate method of calculating the density of states in such
systems for low concentrations of such clusters.

Caption for figure.

Fig.1 (Color on line) The density of states $g(\omega )$ as a function of
the frequency $\omega $ for a system containing a concentration 1.5\% of
defects with six weak springs all having $K_{1}/K_{0}=0.1$. The \ continuous
line shows $g(\omega )$ for the system without defects, the dashed line the
density calculated using equation (17) and the dotted line that obtained
from the simulations.


\begin{thebibliography}{*}
\bibitem[*]{Author} Corresponding author

\bibitem{CPA} P. Soven, Phys. Rev \textbf{156} 809 (1967); B. velicky, Phys.
Rev. \textbf{184} 614 (1969).

\bibitem{Odagaki} T. Odagaki and M. Lax, Phys. Rev. B \textbf{24}, 5284
(1981).

\bibitem{Webman} I. Webman, Phys. Rev. Lett. \textbf{47}, 1496 (1981)

\bibitem{Schir-PRL} W. Schirmacher, G. Diezemann and C Ganter, Phys. Rev.
Lett \ \textbf{81} 136 (1998)

\bibitem{Dunlap} D. H. Dunlap, P. E. Parris and V. M. Kenkre, Phys. Rev.
Lett \textbf{77 }542 (1996)

\bibitem{HCPA} T Odagaki and F. Yonezawa, Solid State Commun.\textbf{27},
1203 (1978); T Odagaki and F. Yonezawa, J. Phys. Soc. Japan \textbf{47}, 379
(1979)

\bibitem{HCPA-obj} J. Van der Rest, Ph. Lambin and F. Brouers, Solid State
Comm. \textbf{38}, 1139. (1981)

\bibitem{Economou} E.N. Economou, \textit{Green's functions in quantum
physics} (Springer, Berlin, 1990)

\bibitem{Schir} W. Schirmacher and G. Diezemann, Ann.Phys.(Leipzig) \textbf{8%
} 727 (1999)
\end{thebibliography}
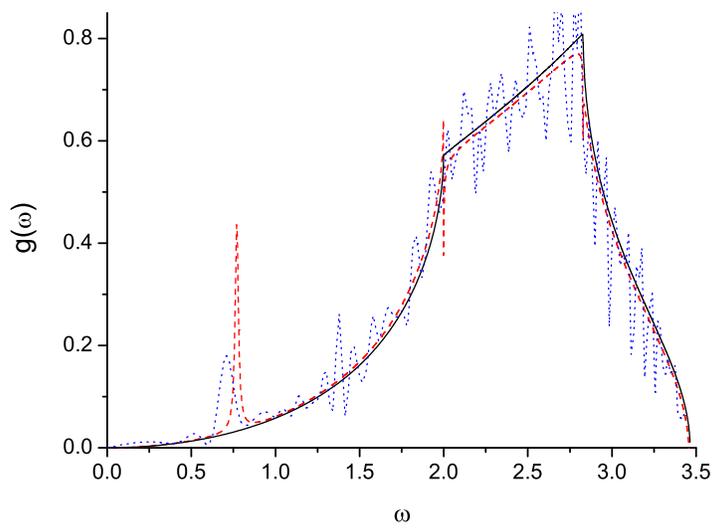
\end{document}